\documentclass[letter]{aa}  

\usepackage{graphicx}
\usepackage{txfonts}
\usepackage{caption}
\usepackage{mathabx}

\begin{document}

\title{Capturing the oxidation of silicon carbide in rocky exoplanetary interiors}

\titlerunning{Oxidation of SiC in exoplanetary interiors}
\authorrunning{K. Hakim et al.}

\author{Kaustubh Hakim\inst{1,2,*} \and Wim van Westrenen\inst{2} \and Carsten Dominik\inst{1} }

\institute{Anton Pannekoek Institute for Astronomy, University of Amsterdam, Science Park 904, 1098 XH Amsterdam, The Netherlands 
        \and Department of Earth Sciences, Vrije Universiteit, De Boelelaan 1085, 1081 HV Amsterdam, The Netherlands}

\date{*Corresponding author email: hakim.kaustubh@gmail.com \\
\\
Received July 25, 2018; Accepted September 26, 2018}

 
  \abstract
   {Theoretical models predict the condensation of silicon carbide around host stars with C/O ratios higher than 0.65 (cf. C/O$_{\mathrm{Sun}}$ = 0.54), in addition to its observations in meteorites, interstellar medium and protoplanetary disks. Consequently, the interiors of rocky exoplanets born from carbon-enriched refractory material are often assumed to contain large amounts of silicon carbide. }
   {Here we aim to investigate the stability of silicon carbide in the interior of carbon-enriched rocky exoplanets and to derive the reaction leading to its transformation.}
   {We performed a high-pressure high-temperature experiment to investigate the reaction between a silicon carbide layer and a layer representative of the bulk composition of a carbon-enriched rocky exoplanet.}
   {We report the reaction leading to oxidation of silicon carbide producing quartz, graphite, and molten iron silicide. Combined with previous studies, we show that in order to stabilize silicon carbide, carbon saturation is not sufficient, and a complete reduction of Fe$^{2+}$ to Fe$^{0}$ in a planetary mantle is required, suggesting that future spectroscopic detection of Fe$^{2+}$ or Fe$^{3+}$ on the surface of rocky exoplanets would imply the absence of silicon carbide in their interiors.}
   {}


   \maketitle
%


\section{Introduction}\label{paper3:introduction}

Silicon carbide grains have been observed in meteorites \citep{Huss2003}, in the interstellar medium \citep{Min2007}, and in protoplanetary disks \citep{Fujiyoshi2015}. Although the Earth is poor in carbon, carbon-rich carbonaceous chondrite meteorites, diamonds discovered in the ureilite parent body \citep{Nabiei2018}, and the proposed presence of graphite on Mercury's surface \citep{Peplowski2016} suggest locally carbon-rich environments in the early solar system. Moreover, chemical simulations of protoplanetary disks around host stars with C/O ratios higher than 0.65 (cf. C/O$_{\mathrm{Sun}}$ = 0.54) result in the condensation of refractory minerals ranging from oxides, silicates, and metals to silicon carbide and graphite \citep{Bond2010b,CarterBond2012b,Moriarty2014}. Dynamical simulations show that these refractory minerals end up in the interiors of rocky exoplanets in different proportions with up to 47~wt\% carbon \citep{CarterBond2012b}. Because of the low density of the carbon-bearing minerals, silicon carbide and graphite are used to explain the low-density rocky exoplanets in the mass-radius diagram with insignificant gas envelopes \citep{Seager2007,Madhusudhan2012}.

Because of speculations of silicon carbide in exoplanetary interiors, physical properties of silicon carbide are being studied extensively at high pressures and temperatures \citep{Wilson2014,Nisr2017,Daviau2017,Miozzi2018}. Significant amounts of silicon carbide in exoplanetary interiors would have a major impact on the thermal evolution and geodynamical processes on such exoplanets because its thermal conductivity is abnormally high \citep{Nisr2017}. There is no question that silicon carbide is highly refractory in nature because its extremely high melting temperatures facilitate its survival in protoplanetary disks once formed. However, the pressures in the interior of planets are orders of magnitude higher than those in protoplanetary disks, which strongly affects its stability. 

Laboratory experiments suggest that SiC is not stable at high-pressure high-temperature conditions resembling those in carbon-enriched exoplanetary interiors \citep{Hakim2018b}. Moissanite (naturally occurring SiC), a rare mineral in the Earth, is known to be unstable in the carbon-poor conditions dominating Earth's mantle and crust, and its formation in Earth is attributed to extremely reducing local conditions \citep{Schmidt2014,Golubkova2016}. Experiments in the Fe-Mg-Si-C-O (FMS+CO) system show that silicon carbide is stable only at extremely low oxygen fugacities of about 6 log units below the iron-w{\"u}stite (IW) buffer ($\log f_{\mathrm{O_{2}}}$=IW$-$6) \citep{Takahashi2013}. At oxygen fugacities above IW$-$6, silicon carbide becomes oxidized, but it is not clear which reaction drives the instability of silicon carbide in a carbon-enriched exoplanetary interior. In this study, we performed an experiment at 1~GPa and 1823~K by juxtaposing an SiC layer and a bulk composition representative of a small carbon-enriched rocky exoplanet \citep{Hakim2018b} (see Table \ref{tab:StartingMaterial}). 

\section{Experimental and analytical method}\label{paper3:method}

\subsection{Starting materials}\label{paper3:methodComposition}

We prepared a mixture of eight major elements (Fe, O, Si, Mg, Al, Ca, S, and C) representative of the bulk composition of a carbon-enriched exoplanet. The proportions of elements in these mixtures are based on the sequential condensation modeling of the protoplanetary disk of HD19994 at 1 astronomical unit (AU) and 0.15 Myr from the study by \citet{Moriarty2014} (see Table \ref{tab:StartingMaterial}). To prepare the chemical mixtures, the starting materials were mixed in proportions shown in Table \ref{tab:StartingMaterial}. In the first step, $\mathrm{SiO_{2}}$ (99.9\% $\mathrm{SiO_{2}}$ powder from Alfa-Aesar), MgO (99.95\% MgO powder from Alfa-Aesar), $\mathrm{Al_{2}O_{3}}$ (99.95\% min alpha $\mathrm{Al_{2}O_{3}}$ powder from Alfa-Aesar), $\mathrm{CaCO_{3}}$ (99.95-100.05\% ACS chelometric standard $\mathrm{CaCO_{3}}$ powder from Alfa-Aesar), and $\mathrm{Fe_{2}O_{3}}$ (99.9\% $\mathrm{Fe_{2}O_{3}}$ powder from Alfa-Aesar) were homogenized with an agate mortar under ethanol. The oxide-carbonate mixture was decarbonated in a box furnace by gradually increasing the temperature from 873~K to 1273~K in six hours. The decarbonated mixture, placed in a Pt-crucible, was first subjected to 1873~K in a box furnace for 30 minutes and then quenched to room temperature by immersing the bottom of the Pt-crucible in water, leading to the formation of glass. The glass was ground to a homogeneous powder using an agate mortar under ethanol. Fe (99.95\% Fe powder, spherical, \textless 10 microns from Alfa-Aesar) and FeS (99.9\% FeS powder from Alfa-Aesar) were then added to the glass powder. In the carbon-enriched case, C (99.9995\% Ultra F purity graphite powder from Alfa-Aesar) was also added to the glass powder. The final mixture was again homogeneously ground with an agate mortar and stored in an oven at 383~K until use. SiC ($-$400 mesh particle size, $\ge$ 97.5\% SiC from Alfa-Aesar) was also ground with an agate mortar and stored separately.

\begin{table}[!h]
\caption{\label{tab:StartingMaterial} Bulk composition of a carbon-enriched planetesimal based on protoplanetary disk evolution modeling around high C/O stars from \citet{Moriarty2014} }
\small
\begin{center}
\begin{tabular}{lr|lr} \hline \hline \rule[0mm]{0mm}{0mm}
Element & mol\% & Material & wt\% \\[1mm]                                                                                                                                       
\hline \\[-1mm]
Si    & 11.4 & $\mathrm{SiO_{2}}$     & 30.1 \\
Mg    & 11.4 & MgO                    & 20.2 \\
O     & 45.8 & FeO$^\dagger$          & 27.3 \\
Fe    & 11.4 & Fe                     &  2.2 \\
S     &  1.9 & FeS                    &  7.0 \\
Al    &  1.4 & $\mathrm{Al_{2}O_{3}}$ &  3.1 \\
Ca    &  0.7 & CaO$^\dagger$          &  1.7 \\
C     & 16.0 & C                      &  8.4 \\[1mm]
\hline
\end{tabular}
\end{center}\caption*{\footnotesize $^\dagger$ CaO and FeO are obtained from $\mathrm{CaCO_{3}}$ and $\mathrm{Fe_{2}O_{3}}$ after decarbonation. The starting composition reflects the composition of the powder after the decarbonation. }
\end{table}

\subsection{High-pressure high-temperature experiments}\label{paper3:methodExperiment}

The experiment was conducted in an end-loaded piston-cylinder apparatus at a pressure of 1~GPa  and temperature of 1823~K in a 12.7 mm (half-inch) sample assembly. Carbon-enriched composition powder was inserted in a 1.6 mm wide graphite capsule, filling the capsule approximately 60 percent by volume. Silicon carbide powder was inserted on top of the carbon-enriched planetary bulk composition, filling the remaining 40 percent by volume of the capsule. The capsule was then sealed with a stepped graphite lid. This graphite capsule was put into a 2 mm wide Pt capsule that was sealed and arc-welded on both ends using a Lampert PUK 3 welder. The Pt capsule was placed in a MgO rod sealed with MgO cement to hold the Pt capsule in place. The MgO rod was introduced in a graphite furnace, thermally insulated by surrounding it with an inner pyrex sleeve and an outer talc sleeve. A four-bore $\mathrm{Al_{2}O_{3}}$ rod through which thermocouple wires were threaded was placed on the top of MgO rod. Pressure calibration of the assembly was performed by bracketing the albite to jadeite plus quartz and fayalite to ferrosilite plus quartz transitions \citep{VanKanParker2011}. The resulting pressure correction of 3\% is consistent with literature data \citep{McDade2002}. A hardened silver steel plug with a pyrophillite ring and a hole for thermocouple were placed on top of the talc-pyrex assembly. A $\mathrm{W_{97}Re_{3}/W{75}Re_{25}}$ (type D) thermocouple was placed in the thermocouple hole directly above the Pt capsule. The distance of 1$-$3.5 mm between the thermocouple tip and the sample produced a temperature difference of $\sim$10~K \citep{Watson2002}. To reduce the porosity of the graphite capsule, the sample assembly was sintered at 1073~K and 1~GPa for 1~h before further heating and pressurization. During heating to run temperature, the pressure was increased continuously using the hot-piston-in technique. The temperature was increased at a rate of 100~K/min. The experiment was run for the duration of 3.5~h and was subsequently quenched to \textless 500~K within $\sim$15 s by switching off the electric power to the heater.

\subsection{Analytical procedure}\label{paper3:methodAnalysis}

The recovered samples were mounted in one-inch diameter mounts using petropoxy resin, cut longitudinally, polished with grit-paper and fine-polished down to a 1/4~$\mathrm{\mu}$m finish. The polished samples were carbon-coated to ensure electrical conductivity of the surface during electron probe micro-analysis (EPMA). Major element contents of the experimental charges were determined using wavelength dispersive spectroscopy on the five-spectrometer JEOL JXA-8530F Hyperprobe Field Emission Electron probe micro-analyzer at the Netherlands National Geological Facility, Utrecht University. For Fe-C and Fe-Si alloys, samples were coated with aluminum instead of carbon, and analyses were performed using a JEOL JXA 8530F Hyperprobe at Rice University, Houston following the analytical protocol \citep{Dasgupta2008}. We used a series of silicate, oxide, and metal standards and conditions of 15 nA beam current and 15 kV accelerating voltage. Analyses were made with a defocused beam to obtain the compositions of the metal (2$-$10~$\mathrm{\mu}$m diameter) and silicate (5$-$20~$\mathrm{\mu}$m diameter) phases. Standards for the quantitative analysis of Mg, Fe, Si, Al, and Ca in silicate minerals were forsterite, hematite, forsterite, corundum, and diopside, respectively. Standards used for measuring Fe, Si, C, S, and O were Fe-metal, Si-metal, experimentally synthesized Fe$_{3}$C, natural troilite, and magnetite, respectively. Counting times were 30~s for Fe (hematite and Fe-metal), Si, C, O, Mg, and Al, and 20~s for Ca and S. Data reduction was performed using the $\Phi$(rZ) correction \citep{Armstrong1995}. The instrument calibration was deemed successful when the composition of secondary standards was reproduced within the error margins defined by the counting statistics.

\section{Results and discussion}\label{paper3:results}

\begin{figure*}[!htp]
  \centering
  \medskip
  \includegraphics[width=.93\textwidth]{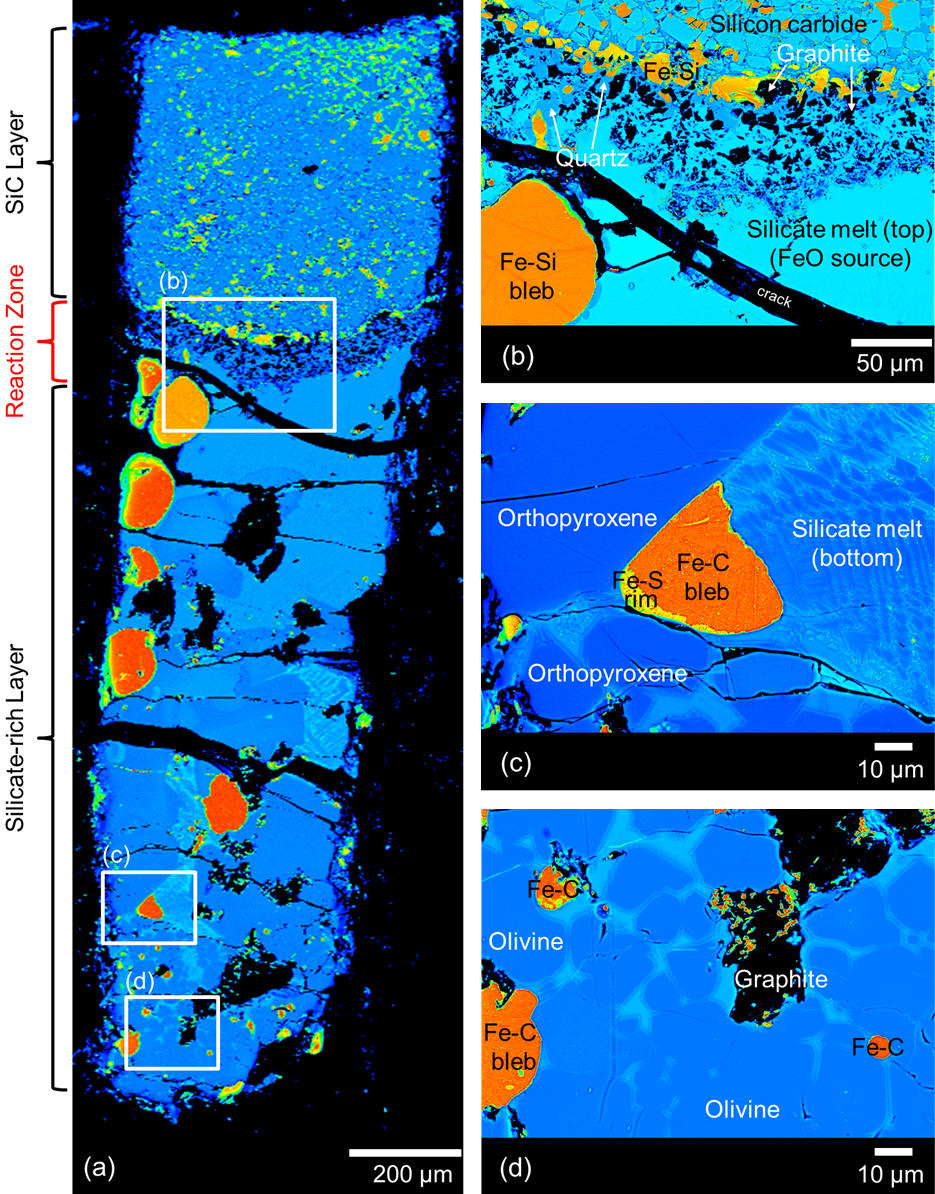}
  \caption[Microprobe Image]{ False-color backscattered electron images of the run product at 1~GPa and 1823~K. (a) A cross-section of the whole capsule is shown. The reaction zone lies between the silicon carbide layer and the silicate-rich layer representing a carbon-enriched rocky exoplanet. (b) The reactants (SiC and FeO from silicate melt) and the products (quartz, graphite, and Fe-Si alloy) are clearly visible. (c) One of the Fe-C blebs has a Fe-S rim, surrounded by orthopyroxene crystals and silicate melt. (d) At the bottom of the capsule, olivine crystals are present instead of orthopyroxene.
        }
  \label{fig:ReactionImage}
\end{figure*} 

The backscattered electron image (Fig. \ref{fig:ReactionImage}a) of the experimental run product shows a clear reaction zone between the SiC layer (top) and the silicate-rich layer representing a carbon-enriched rocky exoplanet (bottom). Tables \ref{tab:PhaseCompositionSilicates} and \ref{tab:PhaseCompositionIronAlloys} give the compositions of the phases analyzed using wavelength-dispersive spectroscopy, whereas graphite was identified using energy-dispersive spectroscopy. The reaction zone contains grains of C (graphite) and SiO$_{2}$ (quartz), and molten Fe-Si alloy (iron silicide) (Fig. \ref{fig:ReactionImage}b). SiC grains become oxidized in the reaction zone and no SiC is present in or below the reaction zone. Although most of Fe-Si alloy moves to the silicate-rich layer, a small portion of it moves up through the silicon carbide layer. The silicate melt pool present below the reaction zone is enriched in SiO$_{2}$ compared to the silicate melt at the bottom of the capsule because of the progressive dissolution of quartz formed in the reaction zone.

\begin{table*}[!htp]
\caption{\label{tab:PhaseCompositionSilicates} Composition of silicate phases} 
\small
\begin{center}
\begin{tabular}{lr|rrrrrrr} \hline \hline \rule[0mm]{0mm}{0mm}
Phase & $n$ & \multicolumn{1}{|c}{$\mathrm{SiO_{2}}$} & \multicolumn{1}{c}{MgO} & \multicolumn{1}{c}{FeO} & \multicolumn{1}{c}{$\mathrm{Al_{2}O_{3}}$} & \multicolumn{1}{c}{CaO} & \multicolumn{1}{c}{$\mathrm{S}$} & \multicolumn{1}{c}{Sum} \\[2mm]                                                                                                                         
\hline \\[-2mm]
Quartz specks          &  4 & 99.64 (0.49) &  0.20 (0.17) & 1.06 (0.17) & 0.15 (0.06) & 0.05 (0.03) & 0.07 (0.05) & 101.15 (0.23) \\
Silicate melt specks   &  9 & 62.86 (0.08) & 19.92 (0.15) & 5.02 (0.05) & 6.44 (0.05) & 4.46 (0.03) & 0.13 (0.01) & 98.84 (0.08) \\
Silicate melt pool     & 20 & 61.81 (0.21) & 20.60 (0.44) & 5.01 (0.15) & 6.77 (0.08) & 4.47 (0.24) & 0.14 (0.00) & 98.79 (0.23) \\
Orthopyroxene          & 10 & 58.31 (0.22) & 36.38 (0.11) & 4.12 (0.15) & 0.56 (0.02) & 0.37 (0.01) & $<$DL & 99.73 (0.12) \\
Olivine (bottom)       & 12 & 41.47 (0.29) & 52.95 (0.26) & 5.48 (0.24) & 0.12 (0.05) & 0.11 (0.02) & $<$DL & 100.13 (0.19) \\
Silicate melt (bottom) &  8 & 53.37 (0.42) & 26.45 (1.63) & 6.16 (0.52) & 8.21 (0.55) & 4.28 (0.88) & 0.10 (0.03) & 98.58 (0.83) \\[2mm]

\hline
\end{tabular}
\end{center}\caption*{\footnotesize \textit{Note:} All compositions are in wt\% with 1$\sigma$ error given in parantheses. $n$ is the number of analytical points. $<$DL implies the measurements were below the detection limit. Pt contamination in all metallic phases was below the detection limit of approximately 0.07 wt\%. At these levels the activity of iron in these metal phases is unaffected \citep[e.g.,][and references therein]{Steenstra2018a}.}
\end{table*}

\begin{table*}[!htp]
\caption{\label{tab:PhaseCompositionIronAlloys} Composition of metallic phases} 
\small
\begin{center}
\begin{tabular}{lr|rrrrrl} \hline \hline \rule[0mm]{0mm}{0mm}
Phase  & $n$ & \multicolumn{1}{c}{Fe} & \multicolumn{1}{c}{Si} & \multicolumn{1}{c}{C} & \multicolumn{1}{c}{S} & \multicolumn{1}{c}{O} & \multicolumn{1}{c}{Sum} \\[2mm]                                                                                                                         
\hline \\[-2mm]
SiC specks   &  7 & 0.38 (0.03) & 73.30 (0.18) & 26.32 (0.09)$^\dagger$ & $<$DL & $-$ & 100 \\
Fe-Si specks &  7 & 78.40 (0.09) & 19.83 (0.29) & 0.94 (0.18)$^\ddagger$ & 0.12 (0.01) & $-$ & 99.29 (0.18) \\
Fe-Si bleb   & 10 & 79.66 (0.10) & 18.82 (0.24) & 1.01 (0.15)$^\ddagger$ & 0.15 (0.01) & $-$ & 99.64 (0.15) \\
Fe-C bleb    & 11 & 93.75 (0.41) & $<$DL        & 4.33 (0.10)          & 0.79 (0.14) & 0.64 (0.07) & 98.87 (0.23)\\
Fe-S rim     &  8 & 65.90 (2.01) & $<$DL        & 1.66 (0.79)         & 31.22 (2.03) & 0.74 (0.59) & 98.78 (1.51) \\[2mm]

\hline
\end{tabular}
\end{center}\caption*{\footnotesize \textit{Note:} All compositions are in wt\% with 1$\sigma$ error given in parantheses. $n$ is the number of analytical points. $<$DL implies the measurements were below the detection limit. $-$ implies lack of measurements. $^\dagger$ Calculated by subtracting the total from 100. $^\ddagger$ C-abundance calculated using a model for C-solubility in Fe-Si from \citet{Steenstra2018b}. }
\end{table*}

Previous experiments by \citet{Hakim2018b} have shown that the equilibrium phases for the bottom silicate-rich layer representing a carbon-enriched rocky exoplanet comprise graphite, olivine, silicate melt, and a S-rich Fe-C-S alloy. In the presence of SiC grains, the silicate phases in the silicate-rich layer become richer in MgO and poorer in FeO compared to compositions formed in the absence of the initial SiC layer. Quartz produced during the process reacts with olivine to produce orthopyroxene. The FeO contents of olivine and silicate melt in the previous study \citep{Hakim2018b} were $34.2\pm1.1$~wt\% and $21.3\pm0.1$~wt\%, respectively, whereas we find the FeO content of olivine, orthopyroxene, and silicate melt to be 4$-$6 wt\%. There is about five times less FeO in the silicate-rich layer of our experiment than its equilibrium state in the absence of the SiC layer. Moreover, the formation of Fe-Si and S-poor Fe-C alloy melts implies that most of the Fe$^{2+}$ initially bonded to oxygen is reduced to Fe$^{0}$ in this reaction. Since the reaction products observed in the reaction zone are quartz, iron silicide melt, and graphite, we report the following reaction consuming silicon carbide:

\begin{equation}\label{eq:SiC}
\mathrm{3\ SiC (s) + 2\ Fe\text{-}O (l) \rightarrow SiO_{2} (s) + 2\ Fe\text{-}Si (l) + 3\ C (s)}
.\end{equation} 

In the equation, Fe-O (l) denotes Fe$^{2+}$ bonded to oxygen in silicate melt, and Fe-Si (l) denotes Fe$^0$ bonded to Si$^0$ in metallic liquid. An important consequence of the reduction of Fe$^{2+}$ is that the molar Mg/(Mg+Fe) of olivine and other silicates ($X_{\mathrm{Mg}}$)       are very high because of the lack of FeO. In our experiment we find $X_{\mathrm{Mg}}\sim0.95$ for olivine and orthopyroxene and $X_{\mathrm{Mg}}\sim0.88$ for the silicate melt, which are significantly higher than the equilibrium state \citep{Hakim2018b}, $X_{\mathrm{Mg}}\sim0.77$ for olivine and $X_{\mathrm{Mg}}\sim0.47$ for the silicate melt. Experiments and theoretical modeling have shown that SiC and Fe-Si alloy can be equilibrated with olivine only when its $X_{\mathrm{Mg}}$ {\textgreater} 0.99 \citep{Schmidt2014,Golubkova2016}. Experiments in carbon-saturated FMS+CO system also find SiC to be stable only when olivines have a very high $X_{\mathrm{Mg}}$ of 0.992 \citep{Takahashi2013}. We conclude that almost all Fe$^{2+}$ should be in its reduced state, Fe$^{0}$, to stabilize SiC even in a carbon-enriched rocky exoplanetary interior.

Since the reaction is still in progress, the conditions at the top of the silicate-rich layer are more reducing than at its bottom. Therefore, Fe-Si alloy is only present within the original SiC layer and at the top of the silicate-rich layer, whereas the remainder of the silicate layer contains Fe$\pm$C$\pm$S alloys only. One Fe-C bleb at the bottom of the capsule is surrounded by a Fe-S alloy showing liquid metal immiscibility because of the lowered local S/Fe ratio, as shown in the previous study by \citet{Hakim2018b} (Fig. \ref{fig:ReactionImage}c). Similarly, orthopyroxene, which is relatively oxygen-poorer than olivine, is found in most of the silicate-rich layer except at the bottom  where olivine is present. We do not find any SiC below the reaction zone, which also contains a large Fe-Si bleb. This suggests that SiC forms at even more reducing conditions than needed for the formation of the Fe-Si alloy, which is in contrast to previous modeling studies in the context of moissanite stability in the Earth's mantle, which suggested that SiC and Fe-Si alloy form together at the same oxygen fugacities \citep{Schmidt2014,Golubkova2016}.

The backscattered electron image (Fig. \ref{fig:ReactionImage}a) of the experimental run product shows a clear reaction zone between the SiC layer (top) and the silicate-rich layer representing a carbon-enriched rocky exoplanet (bottom). Tables \ref{tab:PhaseCompositionSilicates} and \ref{tab:PhaseCompositionIronAlloys} give the compositions of the phases analyzed using wavelength-dispersive spectroscopy, whereas graphite was identified using energy-dispersive spectroscopy. The reaction zone contains grains of C (graphite) and SiO$_{2}$ (quartz), and molten Fe-Si alloy (iron silicide) (Fig. \ref{fig:ReactionImage}b). SiC grains become oxidized in the reaction zone and no SiC is present in or below the reaction zone. Although most of Fe-Si alloy moves to the silicate-rich layer, a small portion of it moves up through the silicon carbide layer. The silicate melt pool present below the reaction zone is enriched in SiO$_{2}$ compared to the silicate melt at the bottom of the capsule because of the progressive dissolution of quartz formed in the reaction zone. 

Our experimental conditions are applicable to the magma ocean stage of Ceres- to Pluto-size exoplanets and planetesimals. Larger rocky exoplanets form from the collision and accretion of such planetesimals. Since silicon carbide is not stable in exoplanetary building blocks and based on the rapid pace of the reaction in our experiment, SiC is expected to completely disappear before the formation of larger exoplanets. There may still be cases where an exoplanetary interior is reducing, for example, through a bombardment of SiC-rich meteorites onto the proto-planet. In such a case, the core of the exoplanet will become larger because FeO in the mantle is reduced to Fe and formation of Fe-Si alloy, which will move to the core. This will enrich the core with Si, and the mantle will be deficient in FeO in this case.

In order for SiC to become a dominant mineral in a rocky exoplanet, our experiment indicates that the conditions should be so reducing that such a planet would already contain a Fe-Si alloy core, contrary to the assumption of a pure Fe core with a SiC-C mantle in previous studies \citep{Madhusudhan2012,Nisr2017}. Our results show that SiC is unstable until the conditions are extremely reducing, allowing only for traces of Fe$^{2+}$, and previous studies \citep{Takahashi2013,Hakim2018b} show that graphite/diamond is the dominant carbon-bearing mineral. Hence, care should be taken when assuming SiC in the interior modeling of a carbon-enriched rocky exoplanet. Assuming no effects of atmosphere or water on the composition of the surface of a rocky exoplanet, if Fe$^{2+}$ or even more oxidizing Fe$^{3+}$ is present in the mantle, it should also be present on the surface. Since the presence of Fe$^{2+}$ or Fe$^{3+}$ implies the absence of SiC, future spectroscopic detections of Fe$^{2+}$ or even more oxidizing Fe$^{3+}$ on the surface of a rocky exoplanet would mean that its interior is devoid of SiC. The conversion of SiC into graphite as well as the presence of graphite in carbon-saturated but not extremely reduced rocky exoplanets would have important consequences for the surface composition and therefore for the habitability of such planets.

\begin{acknowledgements}
      This work is part of the Planetary and Exoplanetary Science Network (PEPSci), funded by the Netherlands Organization for Scientific Research (NWO, Project no. 648.001.005). We are grateful to Sergei Matveev and Tilly Bouten from Utrecht University for their technical assistance during EPMA measurements at Utrecht University. We thank Damanveer Grewal and Rajdeep Dasgupta for performing and facilitating analyses of metals in the EPMA Laboratory at Rice University.
\end{acknowledgements}

%
%

\bibliographystyle{aa}
\bibliography{SiliconCarbideReaction}

\end{document}